\documentclass[%
reprint,
superscriptaddress,
showpacs,
preprintnumbers,
 bibnotes,
 amsmath,amssymb,
 aps,
prl
]{revtex4-1}

\usepackage{graphicx}
\usepackage{dcolumn}
\usepackage{bm}
\usepackage{color}

\begin{document}


\title{Nematic crossover in {BaFe$_2$As$_2$} under uniaxial stress}

\author{Xiao~Ren}
\affiliation{International Center for Quantum Materials, School of Physics, Peking University, Beijing 100871, China}
\author{Lian~Duan}
\affiliation{International Center for Quantum Materials, School of Physics, Peking University, Beijing 100871, China}
\author{Yuwen~Hu}
\affiliation{International Center for Quantum Materials, School of Physics, Peking University, Beijing 100871, China}
\author{Jiarui~Li}
\affiliation{International Center for Quantum Materials, School of Physics, Peking University, Beijing 100871, China}
\author{Rui~Zhang}
\affiliation{Department of Physics and Astronomy, Rice University, Houston, Texas 77005, USA}
\author{Huiqian~Luo}
\affiliation{Beijing National Laboratory for Condensed Matter Physics, Institute of Physics, Chinese Academy of Sciences, Beijing 100190, China}
\author{Pengcheng Dai}
\affiliation{Department of Physics and Astronomy, Rice University, Houston, Texas 77005, USA}
\author{Yuan~Li}
\email[]{yuan.li@pku.edu.cn}
\affiliation{International Center for Quantum Materials, School of Physics, Peking University, Beijing 100871, China}
\affiliation{Collaborative Innovation Center of Quantum Matter, Beijing 100871, China}

\begin{abstract}
Raman scattering can detect spontaneous point-group symmetry breaking without resorting to
single-domain samples. Here we use this technique to study $\mathrm{BaFe_2As_2}$, the parent compound of the ``122'' Fe-based superconductors. We show that an applied compression along the Fe-Fe direction,
which is commonly used to produce untwinned orthorhombic samples, changes the structural
phase transition at temperature $T_{\mathrm{s}}$ into a crossover that spans a considerable temperature range above $T_{\mathrm{s}}$. Even in crystals that are not subject to any applied force, a distribution of substantial residual stress remains, which may explain phenomena that are seemingly indicative of symmetry breaking above $T_{\mathrm{s}}$. Our results are consistent with an onset of spontaneous nematicity only below $T_{\mathrm{s}}$.
\end{abstract}


\maketitle

In the phase diagram of iron-based superconductors, superconductivity is commonly situated near the vanishing point of the phase boundary between tetragonal and orthorhombic crystal structures \cite{JohnstonAdvPhys2010,CanfieldAnnualRevCondMat2010,PaglioneNatPhys2010,BasovNatPhys2011}. This commonality has invoked major research efforts to elucidate the nature of the so-called nematic phase, in which the discrete $C_4$ rotational symmetry is lowered into $C_2$ but the lattice translational symmetry is intact. While the nematic phase is widely believed to be electronically driven, it is currently under debate whether the electrons' spin or orbital/charge degree of freedom is in the ``driver's seat'' \cite{FernandesNatPhys2014}. The fact that the tetragonal-to-orthorhombic structural phase transition is closely accompanied by a stripe antiferromagnetic order in the pnictides supports the spin-driven picture, whereas a pronounced uneven occupation of the Fe $d_{xz}$ and $d_{yz}$ orbitals in the orthorhombic phase \cite{YiPNAS2011,YiNJP2012,ZhangPRB2012,NakayamaPRL2014,WatsonPRB2015} supports the orbital-driven picture. To accommodate the fact that FeSe possesses a nematic phase but no stripe antiferromagnetic order at ambient pressure, alternative scenarios for spin-driven nematicity which do not require long-range magnetic order \cite{Wang2015,Yu2015,Glasbrenner2015,Chubukov2015} have also been proposed.

In order to address the ``driver's seat'' question, it is important to study various susceptibilities in the tetragonal phase \cite{FernandesNatPhys2014}.  A prerequisite for such studies is precise knowledge about the onset temperature of spontaneous nematicity, $T_{\mathrm{nem}}$, and in particular, whether $T_{\mathrm{nem}}$ is simply the orthorhombic structural transition temperature $T_\mathrm{s}$ or even higher. The latter can be the case if spontaneous nematicity first develops electronically at $T_\mathrm{nem}$ with minor (although nonzero) influence on the crystal-lattice ground state, and at a lower temperature $T_\mathrm{s}$ the lattice reaches a tipping point and responses strongly to the nematic electronic structure, bringing a pronounced $C_2$ signature to the entire system \cite{KasaharaNature2012}. In such a scenario, experimental techniques that directly probe the electrons are expected to reveal the nematicity in the temperature range $T_\mathrm{s} < T < T_\mathrm{nem}$.

Indeed, resistivity measurements have revealed a pronounced anisotropy in single crystals that are mechanically detwinned by pressing or pulling on one of the Fe-Fe directions \cite{ChuScience2010,TanatarPRB2010,YingPRL2011}: the resistance along the compressed (elongated) direction becomes larger (smaller). This anisotropy is further found to persist to higher temperatures than $T_\mathrm{s}$, an observation that has been attributed to either electronic nematic susceptibility \cite{ChuScience2010,ChuScience2012}, nematic fluctuations \cite{TanatarPRB2010}, or possible nematic transition or crossover above $T_\mathrm{s}$ \cite{TanatarPRB2010,YingPRL2011}. In accordance with the resistivity anisotropy, spectroscopic evidence for a nematicity-related temperature range above $T_\mathrm{s}$ has also been reported, including anisotropic spin excitations seen by neutron scattering \cite{LuScience2014}, $C_2$-symmetric quasiparticle interference patterns seen by scanning tunneling spectroscopy \cite{RosenthalNatPhys2014}, a gap-like feature seen by point-contact spectroscopy \cite{ArhamJPhys2012}, and a splitting of the Fe $d_{xz}$ and $d_{yz}$ electronic bands seen by photoemission \cite{YiPNAS2011,NakayamaPRL2014}. However, results of transport and thermodynamic measurements \cite{ChuPRB2009,BoehmerPRL2014,LuoPRB2015,Boehmer2015} do not support the existence of a well-defined phase transition at $T > T_\mathrm{s}$, which implies that indications for $T_\mathrm{nem} > T_\mathrm{s}$ might be related to the application of a detwinning force on the crystals. A distinct piece of evidence for $T_\mathrm{nem} > T_\mathrm{s}$ came from magnetic torque measurements which revealed $C_2$ rather than $C_4$ symmetry above $T_\mathrm{s}$ \cite{KasaharaNature2012}: the high sensitivity of the technique allows for measurements on tiny crystals which can be naturally dominated by one nematic domain without intentional detwinning. But still, the possible presence of residual stress in the crystals appears difficult to rule out.

In order to verify the possible existence of spontaneous nematicity at $T > T_\mathrm{s}$, and to elucidate the effect of external as well as residual stress on the phenomena related to $C_4$-symmetry breaking, here we report a systematic study of phonon degeneracy in BaFe$_2$As$_2$ using Raman spectroscopy, with and without applied uniaxial stress. $E_g$ phonons at the Brillouin-zone (BZ) center are doubly degenerate because of the global $C_4$ symmetry, and the lifting of such degeneracy can be used to monitor the lowering of symmetry from $C_4$ to $C_2$. Our method has four major advantages over transport and other spectroscopic measurements: (1) the Raman spectroscopic signature for $C_4$-symmetry breaking is peak splitting, and is unaffected by domain distribution; (2) our detection of lattice dynamics (rather than the equilibrium structure) is in principle also sensitive to nematic fluctuations; (3) the effect of applied force can be separately studied; and (4) the small laser-spot size allows us to gain insight into the local stress distribution. We find that a few MPa of applied uniaxial stress is sufficient to induce a nematic crossover that spans a wide temperature range above $T_{\mathrm{s}}$, and that all our observations in ``free-standing'' samples, some of which could have been interpreted as evidence for $T_\mathrm{nem} >T_\mathrm{s}$, are in fact consistent with a distribution of residual stress on the same order of magnitude inside the crystals.

\begin{figure}
\includegraphics[width=3.375in]{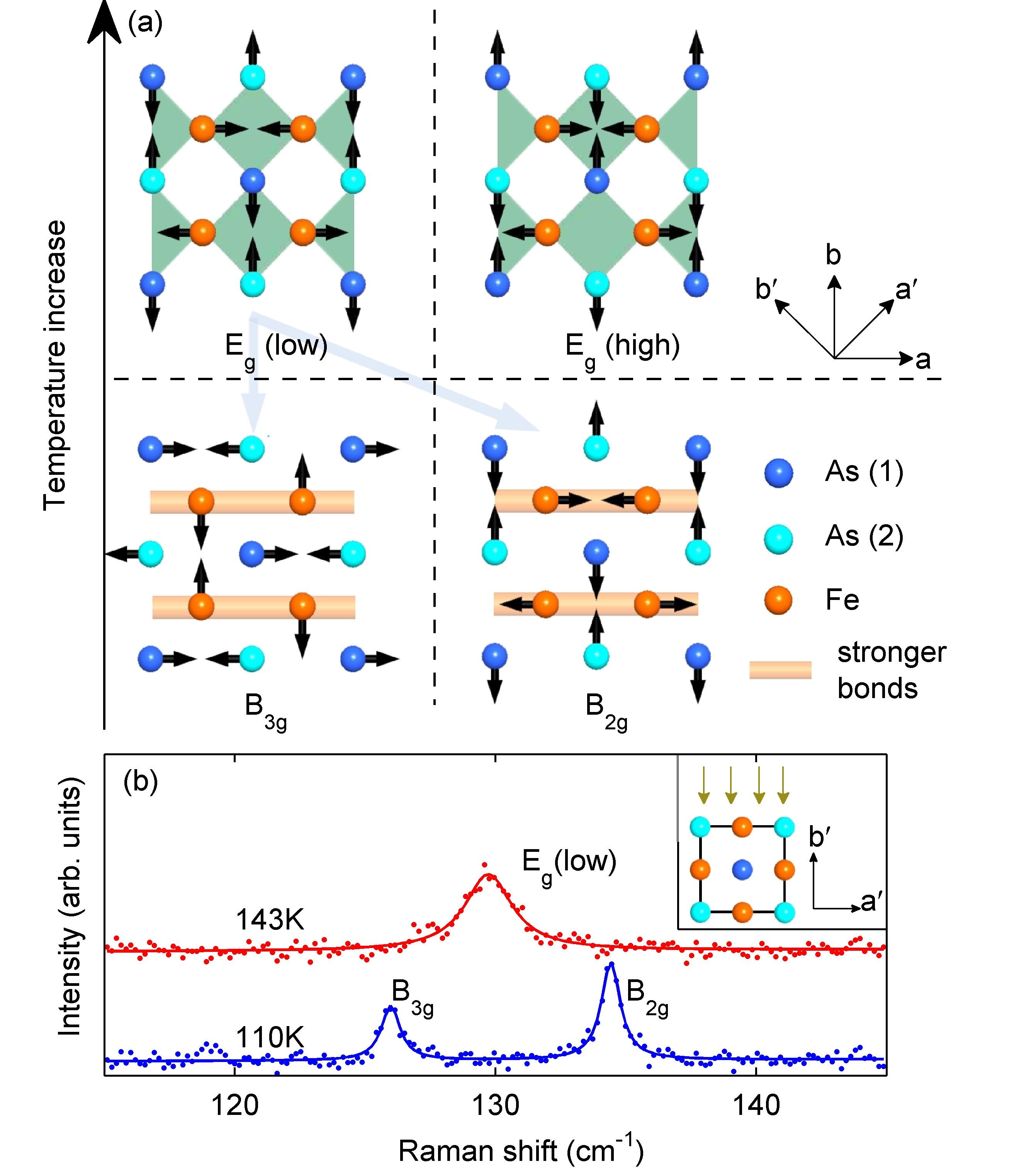}
\caption{\label{Fig1}
(a) Vibrational patterns of two $E_g$ phonons (the 90-degree rotated counterparts are not shown) in the tetragonal phase (upper panels), along with those of $B_{2g}$ and $B_{3g}$ phonons in the orthorhombic phase (lower panels) that derive from the low-energy $E_g$ mode. (b) Raman spectra near the energy of the low-energy $E_g$ mode peak above and below $T_\mathrm{s} = 138$ K. Inset shows the Raman incident-light (arrows) geometry with respect to the crystal lattice.}
\label{Fig1}
\end{figure}

High-quality single crystals of BaFe$_2$As$_2$ were grown by a self-flux method \cite{ChenSupercondSciTech2011}. At high temperature, the crystal structure belongs to the $I4/mmm$ space group and possesses two BZ-center $E_g$ phonon modes, each of which is doubly degenerate and involves displacements of As and Fe atoms parallel to the $ab$-plane. Without performing precise calculations, we schematically show their vibrational patterns in Fig.~\ref{Fig1}(a), from which one can see that the two modes are expected to have rather different energies: the low-energy mode involves half-breathing of Fe-As plaquettes (shaded diamonds) whereas the high-energy mode involves full contraction and expansion of the plaquettes. The two modes have been previously identified by Raman spectroscopy \cite{ChauvierePRB2009} at about 130 cm$^{-1}$ and 270 cm$^{-1}$, respectively. At $T_\mathrm{s}=138$ K, the structure undergoes an orthorhombic transition which is closely accompanied by a magnetic phase transition \cite{RotterPRB2008,KimPRB2011}, and in the orthorhombic phase each of the two $E_g$ modes splits into a $B_{2g}$ (slightly higher energy) and a $B_{3g}$ (slightly lower energy) mode. Since the splitting of the low-energy $E_g$ mode [Fig.~\ref{Fig1}(a)] is more pronounced \cite{ChauvierePRB2009}, here we focus on the study of this mode only.

Our variable-temperature Raman scattering measurements were performed in a confocal back-scattering geometry using a Horiba Jobin Yvon LabRAM HR Evolution spectrometer, equipped with 1800 gr/mm gratings and a liquid-nitrogen-cooled CCD detector. In order to achieve sufficiently high energy resolution, we used the $\lambda= 785$ nm line of a diode laser for excitation, and the laser power was set to 0.9 mW to reduce heating under the focal point which is about 5 microns in diameter. The sample temperature was controlled by a liquid-helium flow cryostat, with the sample kept under better than $5\times10^{-8}$ Torr vacuum at all times. In order to best detect the $E_g$ phonons, one of the incident and scattered light polarizations needs to be parallel to the $c$-axis and the other to the $ab$-plane, which requires measurements to be performed on a surface that is perpendicular to the $ab$-plane, \textit{i.e.}, on the side of the plate-like crystals. High-quality surfaces of this kind were obtained by cleaving the crystals right after freezing them in liquid nitrogen. To obtain high statistical precision, data acquisition for each spectrum took a minimum of twelve hours. A clear splitting of the low-energy $E_g$ peak below $T_\mathrm{s}$ is demonstrated by the data displayed in Fig.~\ref{Fig1}(b).

\begin{figure}
\includegraphics[width=3.375in]{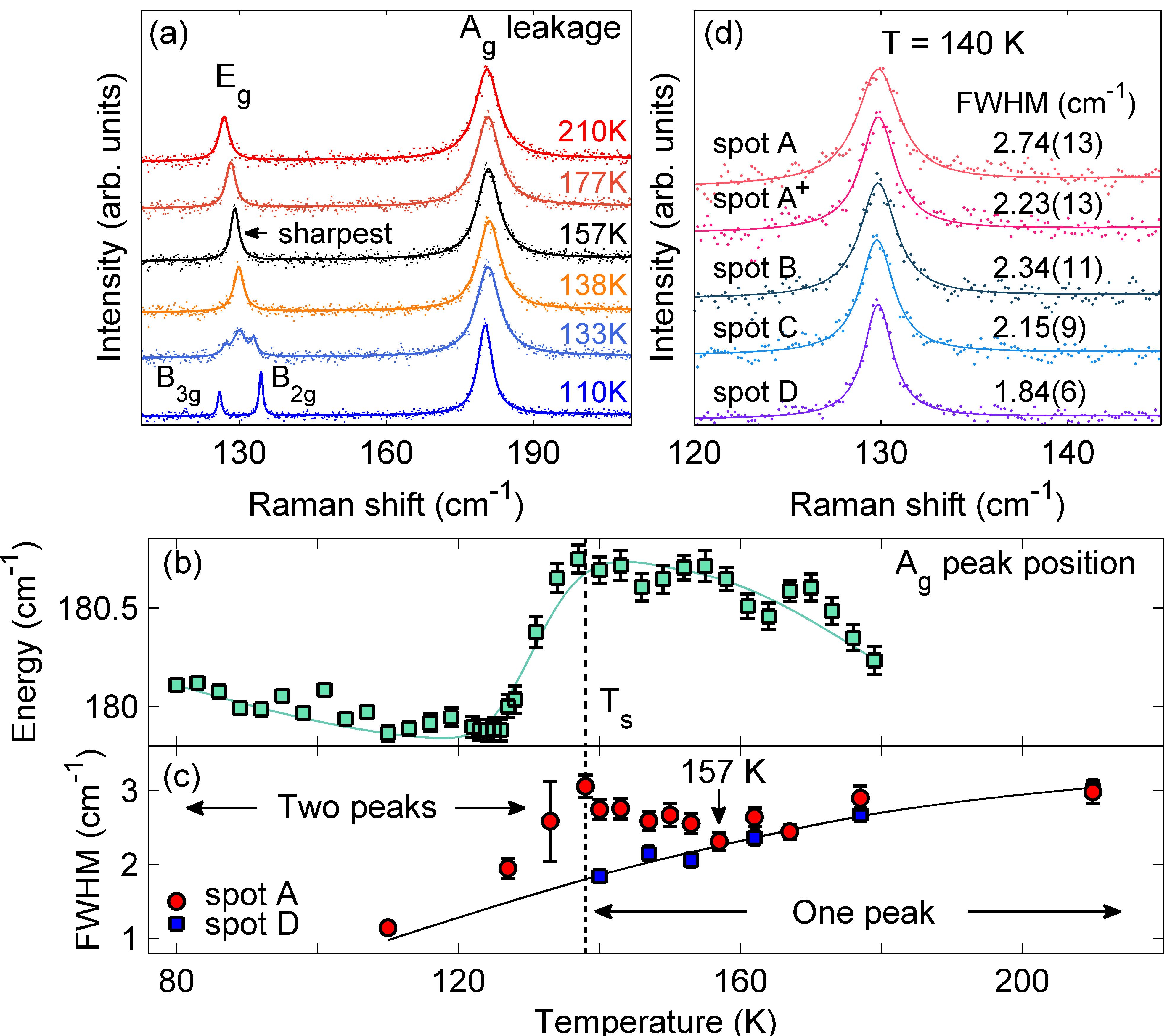}
\caption{\label{Fig2}
(a) Raman spectra measured on a ``free-standing'' sample in $(a^\prime c)$-polarization geometry [see Fig.~\ref{Fig1}(a) for the axis definition] at various temperatures, offset for clarity. (b) $T$ dependence of the $A_g$ phonon energy determined from $(cc)$-polarization spectra obtained with the same laser power. (c) $T$ dependence of the FWHM of the $E_g$ peak measured at two different surface locations specified in (d). The raw spectra were fitted with one Lorentzian peak above $T_{\mathrm{s}}$ (vertical dashed line) and with two peaks below 132 K. Solid lines in (b) and (c) are guide to the eye. (d) Spectra measured in $(a^\prime c)$-polarization geometry at 140 K at four different surface locations, offset for clarity. Spot A$^+$ is the same location as A, but measured after a second cooling.}
\label{Fig2}
\end{figure}

Figure~\ref{Fig2}(a) displays Raman spectra measured at various temperatures with the incident and scattered photon polarizations parallel to the $a^\prime$- [Fig.~\ref{Fig1}(b) inset] and the $c$-directions, which we denote as $(a^\prime c)$-polarization for brevity. The sample was mounted by dipping its bottom into a droplet of viscous Apiezon N-type vacuum grease, which ensured good thermal contact but did not apply any force onto the crystal. We hence refer to this sample as ``free-standing''. Below $T_\mathrm{s}$, the $E_g$ phonon peak splits into $B_{2g}$ and $B_{3g}$ peaks as has already been shown in Fig.~\ref{Fig1}(b), but at 133 K immediately below $T_\mathrm{s}$, a total of three peaks are observed. This can be attributed to laser heating which we estimate to create a temperature distribution of $\sim5$ K above the nominal temperature under the laser spot. Consistently, the $A_g$ phonon energy measured with the same laser power but in $(cc)$ polarization exhibits a pronounced anomaly that is shifted by about $-8$ K from the genuine $T_\mathrm{s}$ [Fig.~\ref{Fig2}(b)].

For regular phonons, their Raman peak widths are expected to decrease with decreasing temperature due to anharmonicity. A close inspection of the spectra above $T_\mathrm{s}$ in Fig.~\ref{Fig2}(a), however, indicates that the smallest full width at half maximum (FWHM) of the $E_g$ phonon peak is realized not immediately above $T_\mathrm{s}$, but around 157 K [``spot A'' data in Fig.~\ref{Fig2}(c)]. It is tempting to interpret the anomalous broadening of the peak between 157 K and $T_\mathrm{s}$ as an indication of onset of spontaneous nematicity above $T_\mathrm{s}$, which might cause the $E_g$ phonon to split into $B_{2g}$ and $B_{3g}$ peaks that are too close in energy to be individually resolved, or as a consequence of anomalous $E_g$ phonon damping brought about by electronic nematic fluctuations. However, such interpretations are not supported by a careful investigation of the phenomenon at different sample-surface locations [Fig.~\ref{Fig2}(d)]: the spectra measured at 140 K at four different locations that are about 50 $\mu$m apart all show different line widths; moreover, measurements performed at one of the same locations (A and A$^+$) but in different cooling cycles result in different line widths beyond our measurement uncertainty. Finally, there is no anomalous broadening of the peak between 157 K and $T_\mathrm{s}$ at spot D, which exhibits the smallest FWHM at 140 K [Fig.~\ref{Fig2}(c)].

\begin{figure}
\includegraphics[width=3.375in]{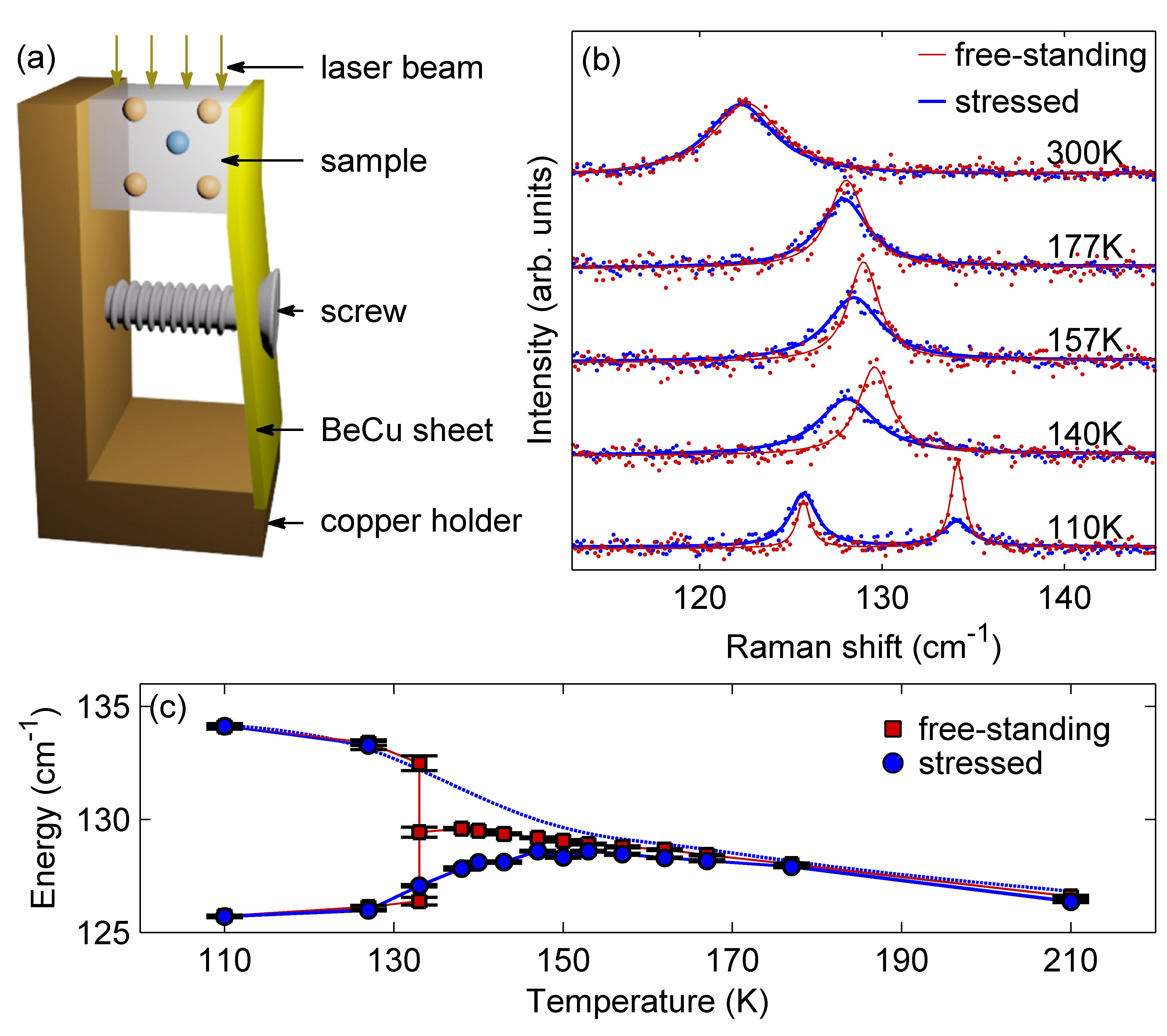}
\caption{\label{Fig3}
(a) A schematic drawing of our sample mount that exerts a uniaxial compression along a Fe-Fe direction onto a crystal. (b) Comparison of Raman spectra at different temperatures measured on stressed and free-standing samples, under $(ac)$- and $(a^\prime c)$-polarization geometries, respectively, vertically offset for clarity. (c) $T$ dependence of phonon energies in stressed and free-standing samples. Solid lines are guide to the eye. Dotted line is an estimate of the $B_{2g}$ phonon energy in the stressed sample, which we are unable to detect in the $(ac)$-polarization geometry.
}
\label{Fig3}
\end{figure}

The above observations suggest an alternative, although relatively mundane, explanation for the anomalous broadening between 157 K and $T_\mathrm{s}$ observed from spot A in Fig.~\ref{Fig2}: the effect could arise from the presence of nonuniform local residual stress, despite the fact that our sample was ``free-standing''. Given the second-order nature of the structural transition at $T_\mathrm{s}$ which is not particularly likely to give rise to residual stress, we speculate that residual stress may be induced by growth defects or our surface-preparation procedure. To explicitly examine the influence of stress, we performed a second set of measurements with the sample mounted in a fashion shown in Fig.~\ref{Fig3}(a). By using an elastic BeCu sheet to press on the crystal, an estimated 5 MPa of uniaxial compressive stress is introduced along one of the Fe-Fe directions, which is within a factor of three comparable to the amount of stress needed for complete sample detwinning \cite{ChuScience2010,LuScience2014}. A comparison of spectra obtained from stressed and free-standing samples is displayed in Fig.~\ref{Fig3}(b). At 110 K, the stressed sample exhibits a pronounced $B_{3g}$ peak and only a weak $B_{2g}$ peak, which is opposite to the free-standing sample. This indicates that we have largely detwinned the crystal, because our measurement of the stressed sample was carried out in the $(ac)$-polarization geometry, and in a fully detwinned sample we should not be able to detect the $B_{2g}$ phonon at all.

According to the data in Fig.~\ref{Fig3}(b), while the uniaxial stress has only minor effect on the peak positions above 160 K or below $T_\mathrm{s}$, a large ``shift'' of the peak is found between 157 K and $T_\mathrm{s}$. Due to the aforementioned reason about our polarization geometry, the peak shifted to lower energy in the stressed sample should be interpreted as one of two split peaks. Assuming that the other peak is symmetrically located on the high-energy side of the original $E_g$ peak, the peak positions versus temperature in both free-standing and stressed samples are displayed in Fig.~\ref{Fig3}(c). Indeed, a rapid splitting of the phonon with cooling is found in the stressed sample below a similar temperature as the onset temperature $T\approx157$ K for anomalous peak broadening in the free-standing sample. For a quantitative comparison, Fig.~\ref{Fig4} displays both the peak shift in the stressed sample and the ``extra'' half width at half maximum (HWHM) of the peak in the free-standing sample, which would correspond to each other if the latter arises from the presence of two unresolvable peaks. In both cases, the data at $T>140$ K can be reasonably well described by Landau's theory, $\Delta E \propto (T - T_\mathrm{s})^{-1}$.

\begin{figure}
\includegraphics[width=3in]{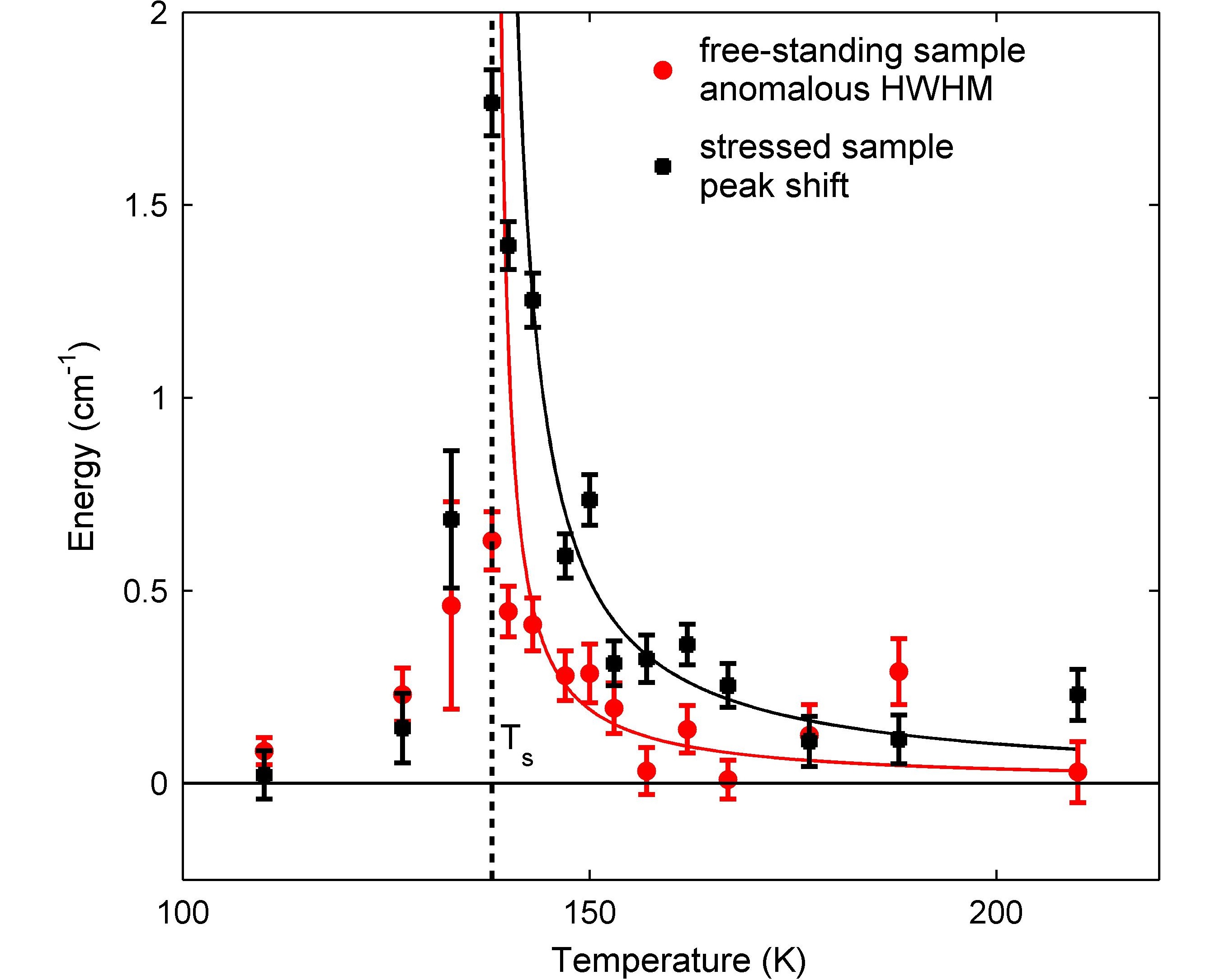}
\caption{\label{Fig4}
Comparison between anomalous peak broadening in free-standing sample (Fig.~\ref{Fig2}) and peak splitting (or shift) in stressed sample (Fig.~\ref{Fig3}). Solid lines are fits according to Landau's theory: $\Delta E \propto (T - T_\mathrm{s})^{-1}$.
}
\label{Fig4}
\end{figure}

The comparison in Fig.~\ref{Fig4} suggests that if the residual stress at spot A in Fig.~\ref{Fig2} amounts to about one quarter that of our applied stress for the measurement in Fig.~\ref{Fig3}, it will be able to explain the anomalous broadening of the $E_g$ phonon below 157 K. Such an amount of residual stress (1-2 MPa) is not unreasonable, as it may arise from processes of thermal cycling, mechanical polishing and cleaving of the crystals, as well as from the presence of growth defects \cite{WithersMST2001}. Together with the nonuniform distribution of $E_g$ peak width at 140 K and the dependence on thermal cycling [Fig.~\ref{Fig2}(d)], our results provide rather strong evidence for a non-negligible influence of residual stress on the phenomena that are seemingly indicative of $C_4$ symmetry breaking above $T_\mathrm{s}$, and are consistent with a lack of genuine phase transition at $T> T_\mathrm{s}$ inferred from transport and thermodynamic measurements \cite{ChuPRB2009,BoehmerPRL2014,LuoPRB2015,Boehmer2015}. Indeed, at sample locations (spot D in Fig.~\ref{Fig2}) where residual stress is negligible, the intrinsic symmetry of the system is $C_4$ above $T_\mathrm{s}$ within our detection limit.

To conclude, our precise Raman scattering measurements of free-standing and stressed BaFe$_2$As$_2$ single crystals show that a moderate uniaxial stress on the order of a few MPa is sufficient to induce appreciable orthorhombicity in the system over a substantial temperature range above $T_\mathrm{s}$. We further show that a hard-to-avoid distribution of residual stress on the order of 1-2 MPa is commonly present even in nominally free-standing single crystals, with typical ``domain'' sizes comparable or larger than the size of our laser spot ($\sim 5$ microns in diameter). Interpretation of measurements done with detwinning forces applied to samples or on small sample volumes, therefore, needs to take these facts into account.

\begin{acknowledgments}

We wish to thank F. Wang, D.-H. Lee, A. B\"ohmer, D.-H. Lu, and Q.-M. Zhang for stimulating discussions. The work at Peking University is supported by the NSF of China (No. 11374024) and the NBRP of China (No. 2013CB921903).

\end{acknowledgments}

\bibliography{Reference}

\end{document}